\documentclass[10pt,letterpaper]{article}
\usepackage{opex3}
\usepackage{amsmath}
\usepackage{cite}
\DeclareMathOperator{\rect}{rect}
\DeclareMathOperator{\sinc}{sinc}

\newcommand{\be}[0]{\begin{equation}}
\newcommand{\ee}[0]{\end{equation}}
\newcommand{\bea}[0]{\begin{eqnarray}}
\newcommand{\eea}[0]{\end{eqnarray}}

\def\njp{ New\ J.\ Phys. }

\begin{document}

\title{Near-perfect sorting of orbital angular momentum and angular position states of light}


\author{Malcolm N. O'Sullivan,$^{*,1}$ Mohammad Mirhosseini,$^{1}$ Mehul Malik,$^{1}$ and Robert W. Boyd$^{1,2}$}

\address{
$^1$The Institute of Optics, University of Rochester. 320 Wilmot BLDG, 275 Hutchison Rd, Rochester NY 14627, USA
\\
$^2$Department of Physics, University of Ottawa, Ottawa, ON K1N 6N5, Canada
\\
$^*$osulliva@optics.rochester.edu
}

\begin{abstract}
We present a novel method for efficient sorting of photons prepared in states of orbital angular momentum (OAM)  and angular position (ANG).  A log-polar optical transform is used in combination with a holographic beam-splitting method to achieve better mode discrimination and reduced cross-talk than reported previously. Simulating this method for 7 modes, we have calculated an improved mutual information of 2.43 bits/photon and 2.29 bits/photon for OAM and ANG modes respectively. In addition, we present preliminary results from an experimental implementation of this technique. This method is expected to have important applications for high-dimensional quantum key distribution systems.
\end{abstract}



\bibliographystyle{osajnl.bst}


\noindent Sorting photons according to their transverse spatial mode is an interesting problem and the subject of ongoing research \cite{Piestun2000}. Perfect sorting occurs when photons prepared in orthogonal input modes are transformed into modes whose intensities do not overlap in space or in time. The ability to accurately and efficiently discriminate the transverse modes of individual photons is vital for applications seeking to encode quantum information using the spatial degree-of-freedom. Sorting methods should aim to minimize detection errors and loss.

Modes with helical phase fronts and no other azimuthal dependence are of particular interest because their helical structure is preserved in propagation through cylindrically symmetric systems, for example in the free-space propagation between an optical transmitter and receiver with circular pupils. It has been long known that modes with azimuthally dependent phase $e^{i\ell\varphi}$ carry $\ell\hbar$ of orbital angular momentum (OAM) \cite{Allen1992, Yao2011}. The complex field of such a mode can be represented by
\begin{equation}
\label{eq:OAMmode}
u_{\ell}(r,\varphi) = R(r) e^{i\ell\varphi},
\end{equation}
where $R(r)$ is an arbitrary function of the radial coordinate $r$ and $\ell$ is an integer. Owing to their high-dimensionality, OAM modes are a prime candidate for free-space quantum key distribution (QKD) systems \cite{Malik2012, Gibson2004, Wang2012,Groblacher2006}. As such, the ability to efficiently sort single photons based on their OAM mode number has become the focus of current research \cite{Leach2004, Berkhout2010, Gruneisen2011}.

Recently, a method for discriminating light beams based on their OAM quantum number $\ell$ has been demonstrated \cite{Berkhout2010,Lavery2012}. Two phase-only holograms are used to optically map polar coordinates $(r,\varphi)$ in the input plane to rectilinear coordinates in the output plane $(x,y)$ via the log-polar mapping $x=a(\varphi \bmod{2\pi})$ and $y=-a\ln (r/b)$, where $a$ and $b$ are scaling constants \cite{Bryngdahl1974,Saito1983}.  The first hologram, the {\it reformatter}, maps the intensities according to the coordinate transformation.  A second hologram, the {\it corrector}, corrects a residual aberration.  Thus, optical waves with helical phase fronts are transformed into tilted plane waves, which can be sorted at the focus of a lens. While this method is substantially more efficient than previous methods for sorting OAM modes, it is still limited to an efficiency of approximately 80\%\cite{Lavery2012}. That is, for a photon with OAM $\ell\hbar$, there exists an approximately 20\% probability of detecting it with OAM $m\hbar,\, m\neq\ell$.

In this work, we show that the technique of Berkhout {\it et al.} \cite{Berkhout2010} can be combined with a holographic beam-splitting technique to sort OAM modes with near unit efficiency. We do this by modifying the two phase-only elements used in Ref. \cite{Berkhout2010} appropriately. In addition, we show that a similar method can be used for efficient sorting of modes complementary and unbiased to the basis consisting of OAM modes with $\lvert\ell\rvert\leq L$. These modes are represented as a superposition of OAM modes
\begin{equation}
\label{eq:angMode}
\theta_j(r,\varphi) = \frac{1}{\sqrt{2L+1}} \sum_{\ell=-L}^{L} u_{\ell}(r,\varphi) e^{-i 2 \pi j\ell/(2L+1)}.
\end{equation}
and are referred to as angular modes (ANG) because of the angular localization of their intensity patterns. The OAM and ANG bases, by being mutually unbiased with respect to one another, guarantee security against eavesdropping in QKD due to the inability of an eavesdropper to detect a photon simultaneously in both bases\cite{Malik2012, Bennett1984}. Thus, for an OAM-based QKD system, an efficient method of sorting photons in the OAM as well as the ANG basis is essential.

To proceed, we consider in detail the action of the log-polar mapping on the OAM modes given by Eq. (\ref{eq:OAMmode}).  Furthermore, without loss of generality, we drop the radial dependence and treat the problem only in the azimuthal dimension. After applying the coordinate transformation, these fields can be expressed as the truncated plane waves
\begin{equation}
\label{eq:fieldTransformed}
U_\ell(x) = e^{i \ell x/a}\, \rect\left(\frac{x}{2\pi a}\right),
\end{equation}
where ${\rm rect}(x) \equiv 1$ for $\lvert x \rvert < 1/2$ and 0 otherwise.  At the focus of a lens with focal length $f$, we can express the field as
\begin{equation}
\label{eq:fieldFocalPlane}
\tilde{U}_\ell(x^\prime) = (2\pi a)\,\sinc\left(\frac{x^\prime-\Delta \ell}{\Delta}\right),
\end{equation}
with $\sinc(x)\equiv\sin(\pi x)/(\pi x)$ and $\Delta = f\lambda/(2\pi a)$. Thus, an input field with OAM index of $\ell$ is focused to a spot centered about $x^\prime=\Delta \ell$. Due to the non-zero width of these spots, there is a significant amount of overlap between the neighboring modes in the output intensity pattern. This shows that even in principle, the OAM sorter cannot perfectly discriminate between adjacent OAM modes\cite{Lavery2012}. 

However, the amount of spatial overlap between the output modes can be reduced. The key is to recognize that the blurring of the spots in the detection plane is caused by the use of only a single angular cycle of \(\varphi\) \cite{Berkhout2010}. To overcome this, we instead periodically map several cycles of $\varphi$ to the points in the output coordinate \(x\). For example, the field at angle $\varphi$ maps to points $x=a(\varphi+ 2\pi m)$ for all integers $m$. One method to accomplish this mapping is to coherently split the field given in Eq.~(\ref{eq:fieldTransformed}) into an array of $N$ orders spaced by $2 \pi a$. This results in the field $U_\ell^\prime(x) = e^{i \ell x/a} \rect(x/(2\pi a N))/\sqrt{N}$ that when focused yields
\begin{equation}
\label{eq:oamModeEnhanced}
\tilde{U}^\prime_\ell(x^\prime) = \frac{2\pi a}{\sqrt{N}} \sinc\left(\frac{x^\prime-\Delta \ell}{\Delta/N}\right).
\end{equation} 
Hence, the resulting overlap becomes negligible when $N$ becomes large enough and perfect discrimination of OAM fields can be realized.


Angular (ANG) modes can also be sorted using a similar approach.  The coordinate transformation acting on mode $\theta_j$ yields
\begin{equation}
\label{eq:angularTransformed}
\Theta_j(x) = \frac{1}{\sqrt{2L+1}} \rect\left(\frac{x}{2\pi a}\right) \delta_L \left(\frac{x-\Delta_\theta j}{a} \right)
\end{equation}
where $\Delta_\theta=2\pi a/(2L+1)$ and $\delta_L(u)=\sum_{\ell=-L}^L e^{i\ell u}$ is the Dirichlet kernel.  The intensity pattern has a peak at $x=\Delta_\theta j$ and its first zero occurs at $x=\Delta_\theta(j\pm 1)$.  Thus, as with the OAM modes, ANG modes can be sorted using position information. However, the spots corresponding to neighboring modes overlap spatially, leading to crosstalk in detection.  

Like the OAM modes, the resolution of the transformed ANG modes is ultimately restricted by their limited spiral spectrum.  Their OAM content can be seen directly from the Fourier transform of Eq.~(\ref{eq:angularTransformed})
\begin{equation}
\label{eq:angModeFocus}
\tilde{\Theta}^\prime_j(k)= \frac{2\pi a}{\sqrt{2L+1}}\sum_{\ell=-L}^L \sinc\left( a k - \ell\right) e^{-i 2\pi j\ell/(2L+1)}.
\end{equation}

Analogous to the technique we used with the OAM modes, we can extend the spatial bandwidth of the modes in such a way that the performance of the sorting improves.  That is, using a lens of focal length $f$, we focus the transformed angular mode, yielding the field $\tilde{\Theta}^\prime_j\left(2\pi x^\prime/\lambda f\right)$. If we now coherently split this field into an array of $N$ orders spaced by $(2L+1) \lambda f/(2\pi a)$ and refocus this field using another lens of focal length $f$, the resulting field will be identical to the field in Eq.~(\ref{eq:angularTransformed}) but instead the sum will contain $N(2L+1)$ terms. When $N=2n+1$ is odd, the field is written as
\begin{equation}
\label{eq:angModeEnhanced}
\Theta_j^\prime(x)=\frac{1}{\sqrt{N(2L+1)}}\rect\left(\frac{x}{2\pi a}\right)\delta_{nL}\left(\frac{x-\Delta_\theta j}{a} \right),
\end{equation}
which has a narrower width than the field in Eq.~(\ref{eq:angularTransformed}) by a factor of $N$. Consequently, arbitrarily low crosstalk can be achieved in the sorting of ANG modes by choosing a sufficiently large $N$. 

 \begin{figure}
                \centering
                \includegraphics[width=11 cm]{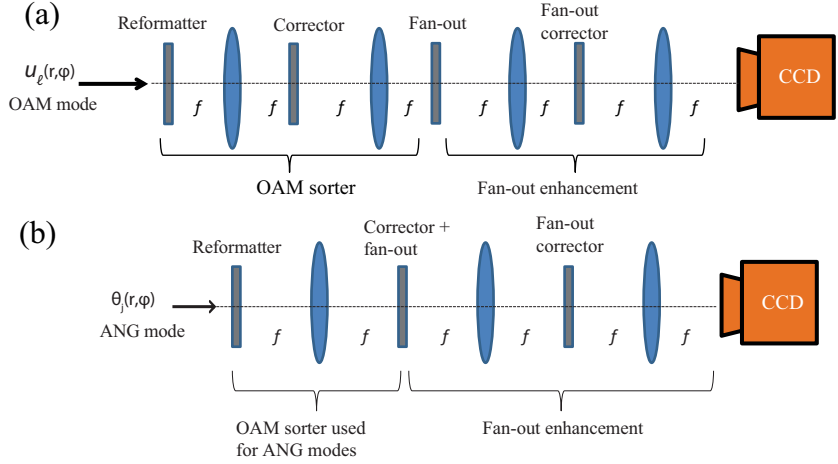}
              \caption{Schematic showing the configuration for enhanced sorting of 1(a) OAM and 1(b) ANG modes. The OAM sorter shown here is from Ref. \cite{Berkhout2010}. The fan-out phase element is combined with the second element of the OAM sorter to create multiple copies of the transformed beam. A phase-correcting element then corrects the relative phases introduced between the copies.}
                \label{fig:setups}
\end{figure}

The implementation of this technique is straightforward.  In addition to optics required for the log-polar coordinate mapping, we require an efficient method of splitting the field into $N$ copies. The fan-out element introduced in Ref. \cite{Prongue1992} is a phase grating designed to diffract an incoming beam into $N$ uniformly spaced orders, each having the same spatial profile and equal energy. For perfect beam splitting, an optical element has to transform an incoming plane wave into a field distribution given by
\begin{equation}
U(x,y)= \sum_{m=1}^{N} A_m e^{i\phi_m}e^{-i 2\pi s_m x/\lambda},
\end{equation}

\noindent where \(A_m\)  is the amplitude, \(\phi_m\) is the phase, and \(s_m\) is the angle of propagation of the \(N\) copies. The fan-out element is the optimal design in the family of phase-only holograms which can approximately achieve this task\cite{Romero2007a}. Generally, the fan-out element introduces a relative phase \(\phi_m\) between the different copies. These are removed with a phase-correcting element in the Fourier plane of the fan-out element. The multiple copies are then Fourier transformed with a lens to a narrower spot than before. Using the specific values of \(A_m\) and \(\phi_m\) given in Ref.\cite{Prongue1992} and \cite{Romero2007a}, we can achieve an efficiency of more than 99\% while splitting the beam to nine copies. 
\begin{figure}
\centering
\includegraphics[width=11cm]{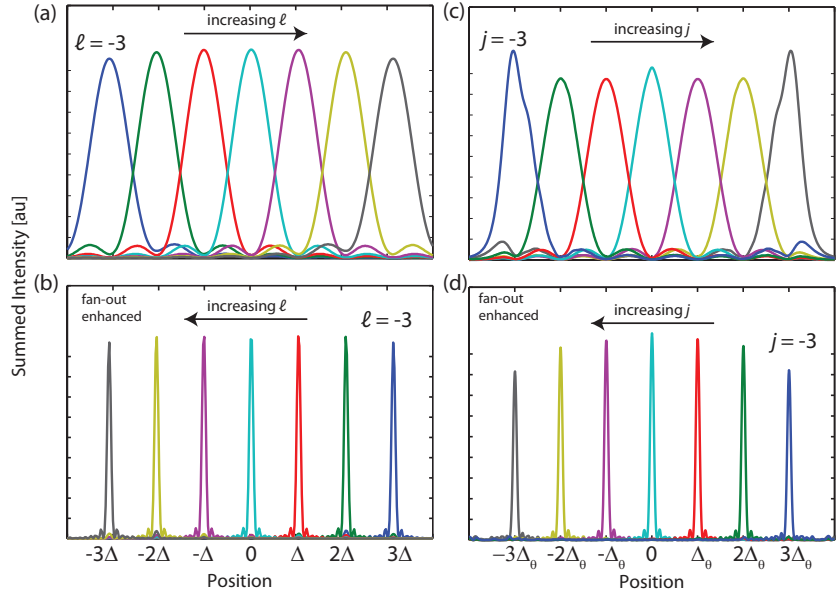}
\caption{Simulation results comparing 2(a) the output from the OAM sorter with 2(b) the output from the \(N=9\) fan-out-enhanced OAM sorter for 7 input OAM modes and comparing 2(c) the output from the ANG sorter with 2(d) the output from the fan-out-enhanced ANG sorter for 7 input ANG modes.  Different colors correspond to different modes. }
\label{fig:OAMANG}
\end{figure}
        

The schematics of the enhanced sorters for the OAM and ANG modes are shown in Fig.~\ref{fig:setups}.  Both use the basic sorter from Ref.~\cite{Berkhout2010}, which performs the log-polar coordinate transformations.  For enhanced sorting of OAM modes, the fan-out element is placed in the Fourier plane of the output of the coordinate mapping optics. For enhanced sorting of ANG modes, the fan-out element is placed directly in the exit plane of the coordinate mapping optics instead.  Both systems use a final phase element to correct the relative phases $\phi_m$ introduced between the multiple copies. A final lens is then used to refocus the light onto the detector. 

 We note that although the OAM sorting method has been described using four distinct phase elements for the sake of clarity, this system can be simplified and requires only two phase elements located in Fourier planes of one another.  For this, the first element would combine the reformatter and fan-out elements while the second element would combine the two corrective elements. This simplified enhanced OAM sorter would perform identically to the four element sorter described previously.  

The enhanced outputs predicted by Eqs. \eqref{eq:oamModeEnhanced} and \eqref{eq:angModeEnhanced} assume ideal implementations of the log-polar mapping and the fan-out operation.  We simulate the fan-out-enhanced sorters using the phase-only implementations identified previously in the text.  Since the input modes are sorted solely according to their position in the $x$ direction, we sum the intensities in the output plane over the $y$ direction. Figures \ref{fig:OAMANG}(a) and (b) show simulation results comparing the output of the OAM sorter from Ref. \cite{Berkhout2010} with the output from the fan-out-enhanced OAM sorter, for different input OAM modes. Figure \ref{fig:OAMANG}(c) and (d) show similar results for the ANG sorter for different input ANG modes. The change from broad, overlapping peaks to narrow peaks with very little overlap clearly shows the dramatic improvement in sorting ability.  The additional change in direction of mode order is a consequence of the fan-out element using two Fourier transforming lenses that together invert the coordinate system.

We quantify this improvement in Table 1 by calculating error percentage and the mutual information for each method. The  error percentage refers to the probability of obtaining mode $x\neq y$ in the output if mode $y$ was input. Mutual information quantifies how much information can be transmitted per photon using given encoding and detection schemes and is given by $ I=-\sum_x P_x\log_2{(P_x)}+\sum_x P_x \sum_y P_{y|x}\log_2(P_{y|x}),$
where the summations are performed over the total number of modes in our system. $P_x$ is the probability that mode $x$ is sent, and $P_{y|x}$ is the conditional probability that mode $y$ is detected given an input mode $x$.  $P_x$ is taken to be uniformly distributed and $P_{y|x}$ is obtained using the intensity patterns shown in the Fig.~\ref{fig:OAMANG} after applying the appropriate normalization.  From Table 1, one can see that the percentage error is substantially reduced in both cases and the mutual information is increased from 1.75 to 2.43 bits/photon for the OAM sorter and from 1.55 to 2.29 bits/photon for the ANG sorter. Thus the fan-out method reduces cross-talk errors and brings the mutual information towards the theoretical limit of $\log_2(n)$.

\renewcommand{\arraystretch}{1.4}

\begin{table*}
\caption{Mutual information (in bits/photon) and error percentages calculated for the OAM and ANG sorters and the fan-out-enhanced (N=9) OAM and ANG sorters when sorting 7 modes. The theoretical limit is shown in the second column and is equal to $\log_2(7)=2.81$.}
{\footnotesize\begin{tabular*}{1.0\textwidth}{@{\extracolsep{\fill}} l c c c c c}

\hline\hline
  Metric\textbackslash Sorter & Limit & OAM (no fan-out) & OAM (fan-out) & ANG (no fan-out)& ANG (fan-out)\\
   \hline
     Mutual Information & 2.81 & 1.75 & 2.43 & 1.55 & 2.29\\
     
  Error Percentage & 0\% & 20\% & 5\% & 23\% & 7\%\\
\hline\hline
\end{tabular*}}

\end{table*}

  \begin{figure}
                \centering
                \includegraphics[width=11 cm]{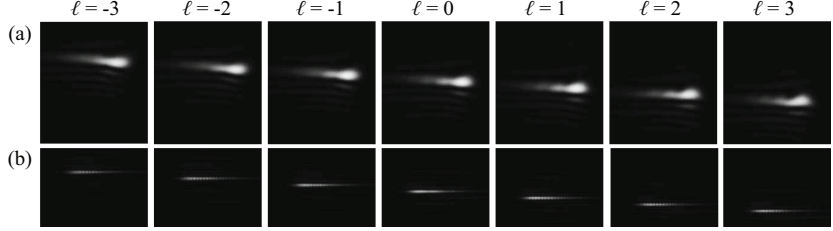}
              \caption{Preliminary results from an experimental implementation of our sorting method. 3(a) Output from the OAM sorter \cite{Lavery2012} for different input OAM modes. 3(b) Output from the fan-out-enhanced OAM sorter proposed in this paper. Here, we used a fan-out element that creates 3 copies of the beam.}
                \label{fig:AllExp}
\end{figure}

Finally, we show preliminary results demonstrating the experimental feasibility of the technique.  We use two refractive elements \cite{Lavery2012} to implement the log-polar mapping and two phase-only spatial light modulators to implement the fan-out operation. Figure \ref{fig:AllExp} shows the CCD images comparing the output from just after the basic OAM sorter to the output after the fan-out enhancement. Here, the fan-out element creates \(N=3\) copies of the incoming beam. One can clearly see the sharping of the output intensity patterns in qualitative agreement with theory. 

In conclusion, we have demonstrated a simple means of combining two existing refractive devices to achieve near perfect sorting of OAM and, for the first time, ANG modes as well.  We have further shown early experimental data demonstrating that this effect can be seen in real-world laboratory conditions.   This work was supported by the DARPA InPho program and the Canadian Excellence Research Chair (CERC) program.
\end{document}